\documentclass[twocolumn,showpacs,superscriptaddress,amsmath,amssymb]{revtex4}
\usepackage{graphicx}
\usepackage{dcolumn}
\usepackage{bm}
\usepackage{color}

\def\bd{
\begin{document}} \def\ed{\end{document}}
\def\bmp{\begin{minipage}} \def\emp{\end{minipage}}
\def\bcc{\begin{center}} \def\ecc{\end{center}}     \def\npg{\newpage}
\def\beq{\begin{equation}} \def\eeq{\end{equation}} \def\hph{\hphantom}
\def\be{\begin{equation}} \def\ee{\end{equation}} \def\r#1{$^{[#1]}$}
\def\n{\noindent} \def\ni{\noindent} \def\pa{\parindent}
\def\hs{\hskip} \def\vs{\vskip} \def\hf{\hfill} \def\ej{\vfill\eject}
\def\cl{\centerline} \def\ob{\obeylines}  \def\ls{\leftskip}
\def\underbar#1{$\setbox0=\hbox{#1} \dp0=1.5pt \mathsurround=0pt
   \underline{\box0}$}   \def\ub{\underbar}    \def\ul{\underline}
\def\f{\left} \def\g{\right} \def\e{{\rm e}} \def\o{\over} \def\d{{\rm d}}
\def\vf{\varphi} \def\pl{\partial} \def\cov{{\rm cov}} \def\ch{{\rm ch}}
\def\la{\langle} \def\ra{\rangle} \def\EE{e$^+$e$^-$} \def\pt{p_{\rm t}}
\def\pti{p_{{\rm t},i}} \def\vti{v_{{\rm t},i}}
\def\ptj{p_{{\rm t},j}}\def\Pt{P_{\rm t}} \def\vt{v_{\rm t}}

\def\bitz{\begin{itemize}} \def\eitz{\end{itemize}}
\def\btbl{\begin{tabular}} \def\etbl{\end{tabular}}
\def\btbb{\begin{tabbing}} \def\etbb{\end{tabbing}}
\def\beqar{\begin{eqnarray}} \def\eeqar{\end{eqnarray}}
\def\\{\hfill\break} \def\dit{\item{-}} \def\i{\item}
\def\bbb{\begin{thebibliography}{9} \itemsep=-1mm}
\def\ebb{\end{thebibliography}} \def\bb{\bibitem}
\def\bpic{\begin{picture}(260,240)} \def\epic{\end{picture}}
\def\akgt{\cl{\bf ACKNOWLEDGMENTS}}
\def\fgn{\noindent{\bf\large\bf figure captions}}
\def\m1{\langle N_p\rangle} \def\u2{\langle N_{\bar p}\rangle} \def\Nap{N_{\bar
p}}
\def\lan{\langle}
\def\ran{\rangle}
\def\p{\pi}
\def\ifmath#1{\relax\ifmmode #1\else $#1$\fi}%
\def\rc{\ifmath{{\mathrm{c}}}}
\def\cut{\ifmath{{\mathrm{cut}}}}
\def\rF{\ifmath{{\mathrm{F}}}}
\def\rK{\ifmath{{\mathrm{K}}}}
\def\rp{\ifmath{{\mathrm{p}}}}
\def\rt{\ifmath{{\mathrm{t}}}}
\def\LAB{\ifmath{{\mathrm{LAB}}}}
\def\cut{\ifmath{{\mathrm{cut}}}}
\def\beq{\begin{equation}}
\def\eeq{\end{equation}}

\newcommand{\cinst}[2]{$^{\mathrm{#1}}$~#2\par}
\newcommand{\crefi}[1]{$^{\mathrm{#1}}$}
\newcommand{\crefii}[2]{$^{\mathrm{#1,#2}}$}
\newcommand{\crefiii}[3]{$^{\mathrm{#1,#2,#3}}$}
\newcommand{\HRule}{\rule{0.5\linewidth}{0.5mm}}

\bd
\title{Azimuthal distributions of radial momentum and velocity
in relativistic heavy ion collisions}

\author{Lin Li} 
\affiliation{Institute of Particle Physics, Hua-Zhong Normal
University, Wuhan 430079, China}
\author{Na Li} 
\affiliation{Hua-zhong University of Science and Technology, 430074,
China}
\author{Yuanfang Wu} 
\affiliation{Institute of Particle Physics,
Hua-Zhong Normal University, Wuhan 430079,
China}\affiliation{Brookhaven National Laboratory, Upton, NY 11973,
U.S.A.}\affiliation{Key Laboratory of Quark $\&$ Lepton Physics
(Huazhong Normal University), Ministry of Education, China }

\begin{abstract}
Azimuthal distributions of radial (transverse) momentum, mean radial
momentum, and mean radial velocity of final state particles are
suggested for relativistic heavy ion collisions. Using transport
model AMPT with string melting, these distributions for Au + Au
collisions at 200 GeV are presented and studied. It is demonstrated
that the distribution of total radial momentum is more sensitive to
the anisotropic expansion, as the anisotropies of final state
particles and their associated transverse momentums are both counted
in the measure. The  mean radial velocity distribution is compared
with the radial flow velocity. The thermal motion contributes an
isotropic constant to mean radial velocity.
\end{abstract}


\maketitle
\section{Introduction}

One of the main goals of current relativistic heavy ion collisions
is to understand the properties of a new formed matter---the
quark-gluon plasma (QGP)~\cite{lab1}. It is well known that one
important character of the formed matter is anisotropic collective
flow. In non-central collisions, the overlap area of two incident
nuclei is an almond shape in the transverse coordinate
plane~\cite{lab2}. This initial geometric asymmetry leads larger
density gradient along the short axis. It in turn pushes the formed
system to expand anisotropically, i.e., large collective flow
velocity in short side direction, which is perpendicular to the
anisotropy in coordinate space. Therefore, the measure of
anisotropic distribution of final state particles should provide
valuable information of the system evolution~\cite{lab2,lab3}.

Conventionally, the azimuthal distribution of multiplicity of final
state particles is presented. Its anisotropy is quantified by the
coefficients of Fourier expansion of the
distribution~\cite{Voloshin}
\begin{equation}
\label{eq1} \frac{dN}{d\phi}\propto \
1+\sum_{n=1}^{\infty}2v_{n}(N)\cos(n\phi),
\end{equation}
where \noindent $\phi$ is the azimuthal angle between the transverse
momentum of the particle and the reaction plane. The Fourier
coefficients is evaluated by,
\begin{equation}
\label{eq2}  v_{n}(N)=\langle\cos(n\phi)\rangle,
\end{equation}
\noindent where $\langle$\ldots$\rangle$ is an average over all
particles in all events, and $v_n(N)$ refers to the anisotropy
coefficient of azimuthal multiplicity distribution. The second
harmonic coefficient $v_{2}(N)$ is the so-called elliptic flow
parameter. It presents the anisotropy of the colliding system and
has the biggest ellipticity at high energy heavy ion
collisions~\cite{b-flow,Danielewicz}. Besides,the azimuthal
asymmetry distribution of energy loss and its coefficients of
Fourier expansion are also studied~\cite{Gyulassya}.


However, the multiplicity distribution only counts the number of
particles emission in a certain azimuthal angle. The expansion of
the system results in not only the anisotropy of multiplicity
distribution but also their associate radial (transverse) momentum.
The total radial momenta at a given azimuthal angle is the
combination of them. Therefore, the azimuthal distribution of radial
momentum should be a more sensitive measure of the anisotropic
expansion, which has not been directly explored before.


In addition to the radial momentum, the radial flow velocity is
another interesting and important quantity. It directly relates to
the equation of state \cite{collectivity} and shear viscous
interactions. For an ideal flow, the radial flow velocity is
isotropy. While, if there are shear interactions, the radial flow
velocity will be different from layer to layer. In hydrodynamics,
the shear viscous interactions are supposed to be proportional to
the gradient of flow velocity~\cite{Landau-book}. The proportional
constant is defined as shear viscosity. The gradient of radial flow
velocity along the azimuthal direction is directly related to the
shear viscous interactions.

Theoretically, the radial flow velocity is a parameter in model
calculations.  It is usually obtained by fitting the spectrum of
transverse momentum~~\cite{PLB503}. Recently, it is further
suggested to extract the radial flow velocity from photon and
dilepton spectrum~\cite{spectra}.

Experimentally, only the radial velocity of final state particles
($\vec{v}$) is measurable. It should be a combination of  velocities
of flow ($\vec{v}_{\rm flow}$) and the random thermal motion
($\vec{v}_{\rm th}$)~\cite{ollitrault}. How to abstract the random
thermal motion from the radial velocity of final state particles and
get the radial flow velocity are not clear. This is why the radial
velocity of final state particles has not been explored in a long
period. It is interesting to see how the radial velocity of final
state particle relates to the radial flow velocity. Therefore, we
further suggest the measurement of the azimuthal distribution of
mean radial velocity of final state particles.

In the second session of the paper, we will give the definitions of
suggested azimuthal distributions of radial momentum and velocity of
final state particles, and the corresponding anisotropic parameters.
In the third session, using the samples generated by AMPT with
string melting model, we show the azimuthal distributions of radial
momentum and the centrality dependence of its anisotropic
parameters. The results are compared with those of the corresponding
azimuthal multiplicity distribution. In the fourth session, the
azimuthal distributions of mean radial velocity at different
centralities are presented, and compared with those given by the
anisotropic blast-wave model~\cite{ko-blastwave,blastwave,heinz}.
Finally, the summary and conclusions are presented.

\section{Azimuthal distributions of radial momentum and velocity}

As indicated, the initial anisotropy in coordinate space in
non-central collisions makes the formed system expand in a
perpendicular almond shape in momentum space. The final state
particles move outward anisotropically. Both the particle density
and the associated momentum behaves anisotropically during the
expansion. The distribution of total transverse momentum at the
different azimuthal directions should be a good measurement for both
of these two effects. The total transverse momentum in the $m$th
azimuthal bin can be defined as
\begin{equation}
\label{eq3}\la\Pt(\phi_m)\ra=\frac{1}{N_{\mathrm{event}}}\sum_{j=1}^{N_{\mathrm{event}}}\left(\sum_{i=1}^{N_{m}}\pti(\phi_m)\right).
\end{equation}
where \noindent $\pti$ is the transverse momentum of the $i$th
particle, $N_{m}$ is the total number of particles, and
$\la\dots\ra$ denotes the average over all events.

In order to see the contributions of radial momentum in particular,
the mean radial momentum in the $m$th azimuthal bin can be defined
accordingly as,
\begin{equation}
\label{eq3} \la\la
\pt(\phi_m)\ra\ra=\frac{1}{N_{\mathrm{event}}}\sum_{j=1}^{N_{\mathrm{event}}}\left(\frac{1}{N_m}\sum_{i=1}^{N_{m}}\pti(\phi_m)\right).
\end{equation}
\noindent Here, the averages $\la\la\dots\ra\ra$ are over all
particles in $m$th angle bin and all events. It records only the
contributions from the transverse momentum of final particles, the
multiplicity effect is canceled by the average over all particles.

The anisotropic parameters of all those azimuthal distributions can
be directly obtained from their Fourier expansions, respectively,
\begin{equation}\label{v2-Pt}
 \frac{d\la\Pt\ra}{d\phi}\propto \
1+\sum_{n=1}^{\infty}2v_{n}(\la\Pt\ra)\cos(n\phi),
\end{equation}
\noindent and
\begin{equation}\label{v2-apt}
 \frac{d\la\la\pt\ra\ra}{d\phi}\propto \
1+\sum_{n=1}^{\infty}2v_{n}(\la\la\pt\ra\ra)\cos(n\phi).
\end{equation}
\noindent $\frac{d\la\Pt\ra}{d\phi}$ and
$\frac{d\la\la\pt\ra\ra}{d\phi}$ are the azimuthal distribution
functions of total radial momentum and mean radial momentum.
$v_{n}(\la\Pt\ra)$ and $v_{n}(\la\la\pt\ra\ra)$ are their
anisotropic parameters, respectively.

Considering the relativistic effect, the transverse (radial)
velocity of the $i$th particle can be written as,
\begin{equation}\label{vti}
v_{t,i}=\frac{p_{t,i}}{m_{\rm
t}}=\frac{p_{t,i}}{\sqrt{m_{0,i}^2+p_{t,i}^2}},
\end{equation}
\noindent where $p_{t,i}$ and $m_{t,i}$ are the transverse momentum
and mass of the $i$th particle, respectively. $m_{0,i}$ is the mass
of $i$th particle in rest frame. The radial velocity fluctuates from
particle to particle. In a given azimuthal direction, the mean
radial velocity can be considered as a good approximation.
Analogously, the azimuthal distribution of mean radial velocity can
be defined as
\begin{equation}\label{vt}
\la\la\vt(\phi_m)\ra\ra=\frac{1}{N_{\mathrm{event}}}\sum_{j=1}^{N_{\mathrm{event}}}\left(\frac{1}{N_{m}}\sum_{i=1}^{N_{m}}\vti(\phi_m)\right),
\end{equation}
\noindent Here, the average is over all the particles in $m$th bin
and events.

The behavior of those suggested observables should provide more
information in anisotropic expansion. In the following, as a
demonstration, we use the generated sample of AMPT with string
melting~\cite{ampt1,ampt2}.  A partonic phase is implemented in the
model and the elliptic flow data from RHIC are well reproduced by
the model~\cite{ampt-v2}. For Au+Au at $\sqrt{s_{NN}}=200$ GeV,
about 1.6 millions minimum bias events are generated.

\section{Azimuthal distributions of radial momentum in AMPT model}

The azimuthal distributions of radial momentum, mean radial
momentum, and multiplicity are presented in Fig.~1(a), (b) and (c),
respectively. Error is statistical only and smaller than the size of
the points. The particles within rapidity range $y\in[-5,5]$ are
counted. These cases are kept in all the following figures.


\begin{figure}
\includegraphics[width=3.4in]{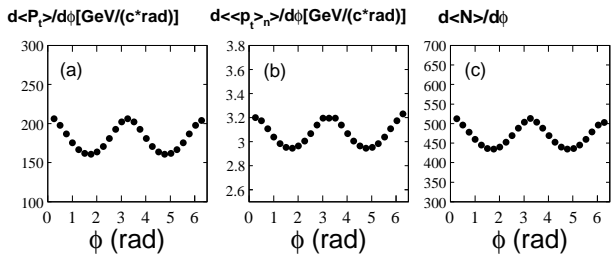}
\caption{\label{Fig. 1} The azimuthal distributions of radial
momentum (a), mean radial momentum (b), and multiplicity (c) for the
sample of Au+Au collisions at $\sqrt{s_{NN}}=200$ GeV generated by
AMPT with string melting.}
\end{figure}

We can see from Fig.~1 that all the observables as a function of
azimuthal angle show the anisotropic shape, $\cos(2\phi)$. It is the
same as multiplicity distribution, the biggest anisotropy of mean
radial momentum distribution appears in in-plan direction as shown
in Fig.~1(b). It indicates that not only the particle density, but
also the associated $\pt$ are larger in in-plane direction. It is
interesting to see if the data at RHIC show the same character as
the model.


\begin{figure}
\includegraphics[width=2.4in]{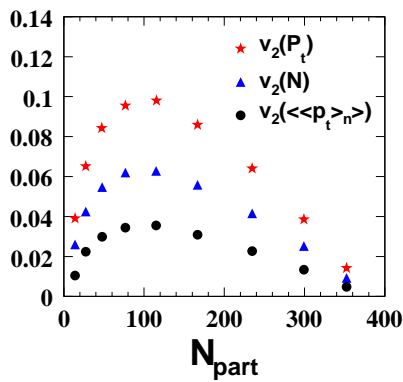}
\caption{\label{Fig. 2}(Color online)The centrality dependence of
elliptic flow parameters deduced from azimuthal distributions of
radial momentum (solid red stars),
 mean radial momentum (solid black cycles), and multiplicity (solid blue
 triangles) for the sample of Au+Au collisions at $\sqrt{s_{NN}}=200$ GeV
 generated by AMPT with string melting.}
\end{figure}

In order to compare the anisotropy effects of these three
distributions qualitatively, the centrality dependence of the
corresponding anisotropic parameter, $v_2$, is presented in Fig.~2.
The anisotropy parameter $v_2$ from different measurements shows the
similar centrality dependencies. At each centralities, the
anisotropy parameter of multiplicity distribution, $v_2(N)$, is
larger than that of the mean radial momentum distribution
$v_2(\la\la p_t\ra\ra)$. The anisotropy parameter of radial
momentum, $v_2(\la P_t \ra)$, is the largest one among the three
variables. It confirms that the anisotropy of radial momentum
distribution including the contributions from a number of particles
and their associated transverse momentum. Therefore, the azimuthal
distribution of radial momentum gives a full count of anisotropic
expansion.

As we know, the anisotropy parameters $v_2$ also depend on $\pt$,
and it increases with $\pt$ when $\pt<2$ GeV/$c$~\cite{star-pt-v2}.
The $\pt$ dependence of the anisotropy parameters of radial momentum
and multiplicity distributions are presented in Fig.~3. The
anisotropy parameter increases with $\pt$ when $ \pt<2$ GeV/$c$, the
same as the data shown. We can also see that the $v_2$ slightly
decreases with $\pt$ when $\pt>2$ GeV/$c$, and it may be contributed
by the hard components~\cite{Rudy}. At a fixed $\pt$ bin, the
anisotropy of the radial momentum is almost the same as that of
multiplicity. This is because the $\pt$ of all particles in a small
given $\pt$ bin are almost the same. The anisotropy of radial
momentum is dominated by that of multiplicity.

\begin{figure}
\includegraphics[width=2.4in]{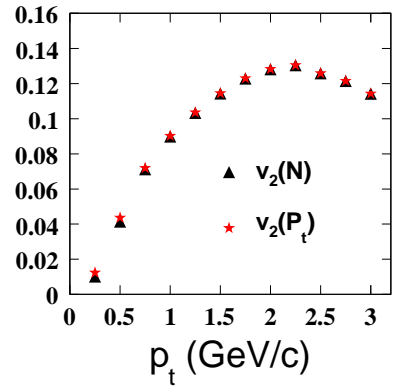}
 \caption{\label{Fig. 3}(Color online) $\pt$ dependence of
 anisotropic parameter of azimuthal distributions of radial momentum (red solid stars),
 and multiplicity (black triangles) for the sample of Au+Au collisions at $\sqrt{s_{NN}}=200$ GeV
 generated by AMPT with string melting.}
\end{figure}

\section{Azimuthal distributions of radial velocity in AMPT model }

The azimuthal distribution of mean radial velocity is presented in
Fig.~4(a). It is a period function and can be well fitted by
\beqar\label{vt-data}\la\la V_t\ra\ra=V_0+V_a\cos(2\phi).\eeqar
\noindent It is the same mode as the flow velocity,
\beqar\label{ko-blastwave-vt}\beta=\beta_0 +
\beta_a\cos(2\phi),\eeqar \noindent which is usually assumed in
blast-wave model in counting the anisotropic
expansion~\cite{ko-blastwave,heinz}.

\begin{figure}
\includegraphics[width=3.5in]{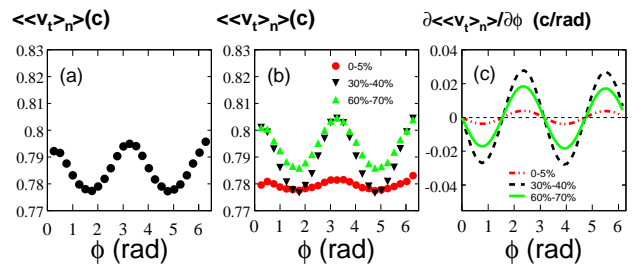}
\caption{\label{Fig. 4}(Color online) The azimuthal distributions of
mean radial velocities of minimum bias sample (a) and the samples of
three different centralities (b), and the azimuthal gradients of (b)
in (c).}
\end{figure}


In order to see the contribution of the random thermal motion, the
mean radial velocity of final state particles in three typical
centralities are presented in Fig.~4(b). In mid-central
($30\%-40\%$) and peripheral ($60\%-70\%$) collisions,  the mean
radial velocities are anisotropy, while it becomes approximately an
isotropy constant in central ($0-5\%$) collisions. This suggests
that the interactions between azimuthal layers are negligible in
central collisions, which is consistent with the expectations of
viscous hydrodynamics~\cite{Heinz-song,Song}. It also shows that the
thermal motions only contribute an isotropic constant to the mean
radial velocity.


As we know, for a system with a fixed temperature, the lighter
particle has higher thermal velocity. In order to test if the $V_0$
is mainly caused by thermal motion, the mean radial velocities of
three different particles and their corresponding fitting parameters
are presented in Fig.~5. Indeed, the lightest pion has the highest
$V_0$, while the heaviest proton has the lowest one.

To see the anisotropy effect alone, we can calculate the gradient of
mean radial velocity along the azimuthal direction. In the case, the
constant part of the mean radial velocity is canceled. Fig.~4(c)
shows the corresponding gradients of Fig.~4(b). In central
collisions, it is approximately zero. The amplitudes in mid-central
collisions are larger than those in peripheral collisions. These
results show that there is almost no gradient of mean radial
velocity in central collisions and it becomes the largest in
mid-central collisions.

Conventionally, the parameters of flow velocity
Eq.~(\ref{ko-blastwave-vt}), $\beta_0$ and $\beta_a$, are obtained
by fitting the spectra of produced particles. Here, we choose the
spectra of pion, proton and kaon from AMPT string melting and get,
$\beta=0.35+0.04cos(2\phi)$. Due to thermal motion, the $\beta_0$ is
not directly comparable with $V_0$. However, $V_a\sim 0.01$ from
corresponding mean radial velocity may be a good approximation of
flow velocity estimated by blast-wave model, where $\beta_a\sim
0.04$.

Certainly, the flow velocities obtained from directly measured
radial velocity and from the spectrum fitting based on blast-wave
model should be better compared by experimental data sample, where
the spectrum is precisely presented. The comparison of these two
methods will lead to a better understanding of the flow velocity.

\begin{figure}
\includegraphics[width=2.5in]{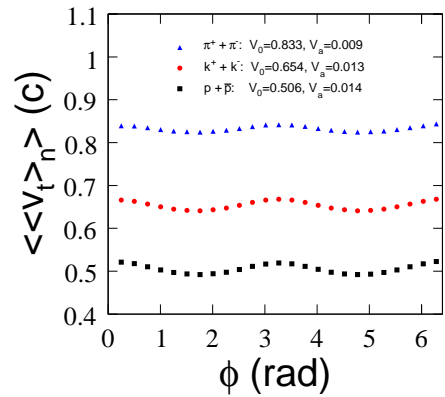}
\caption{\label{Fig. 5}(Color online)The azimuthal distributions of
mean radial velocity of charged pion, charged kaon and (anti)proton.
The lines are fitted by Eq.~(8), and the corresponding fitted
parameters are listed. The errors of the parameters are less than
1\% relative values.}
\end{figure}

\section{Summary and conclusions}

In the paper, we suggest the studies for azimuthal distributions of
radial momentum, mean radial momentum, and mean radial velocity in
relativistic heavy heavy ion collisions.

Using the sample of Au + Au collisions at $\sqrt{s_{NN}}=200$ GeV
produced by a multiphase transport model (AMPT), we find that the
azimuthal distribution of radial transverse momentum indeed counts
the anisotropy of final state particles and their associated
transverse momenta. Thus it presents a full description of
anisotropic expansion at various centralities. Only in small $\pt$
bin, the azimuthal distribution of radial momentum shows the same
anisotropy as that of the multiplicity distribution.

The azimuthal distribution of mean radial velocity is shown to be
the same mode as flow velocity which is usually assumed in
generalized blast-wave model. Its centrality dependency indicates
that thermal motion only contributes an isotropic constant to mean
radial velocity. Its particle mass dependency further shows that the
mass ordering of isotropic mean radial velocity is the same as
thermal motion. The anisotropic mean radial velocity is approximated
to flow velocity, which is obtained from fitting the spectrum of
corresponding particles based on Blastwave model.

Therefore, it is interesting to measure the azimuthal distributions
of radial momentum, mean radial momentum, and mean radial velocity
in current relativistic heavy ion collisions.

The first and last authors are grateful for the valuable comments by
Dr. Zhangbu Xu. The last author is grateful for the hospitality of
BNL STAR group. This work was supported in part by the National
Natural Science Foundation of China under Grant No. 10835005 and MOE
of China under Grant No. IRT0624 and B08033.

\ed
\begin{thebibliography}{99}

\bibitem{lab1} B.M\"{u}ller, Nucl. Phys. A, 2006, 774 : 433

\bibitem{lab2} H. Sorge, Phys. Lett. B, 1997, 402:251; Phys. Rev. Lett.
1997, 78: 2309; Phys. Rev. Lett., 1999, 82: 2048

\bibitem{lab3} P. Danielewicz, Phys. Rev. C, 1995, 51:716


\bibitem{Voloshin} S. Voloshin, Y. Zhang, Z. Phys. C , 1996, 70: 665

\bibitem{b-flow}Sergei A. Voloshin, Arthur M. Poskanzer, and Raimond Snellings,
arXiv:0809.2949

\bibitem{Danielewicz} P. Danielewicz and G. Odyniec, Phys. Lett., 1985, 157B:146

\bibitem{Gyulassya} Miklos Gyulassya, Ivan Viteva,  Xin-Nian Wang et al  Phys.Lett.B , 2002, 526:301-308

\bibitem{collectivity} P. Huovinen , P. V. Ruuskanen, Ann. Rev. Nucl. Part. Sci., 2006,  56
:163; D. A. Teaney, arXiv: 0905.2433

\bibitem{Landau-book} L.D.Landau, E.M. Lifschitz, Fluid Mechanics, Institute of Physical Problems,
U.S.S.R. Academy of Sciences, Volume 6, Course of Theoretical
Physics

\bibitem{PLB503}  Y. Oh, Z. Lin, and  C. M. Ko, arXiv: 0910.1977

\bibitem{spectra} Jajati K. Nayak , Jan-e Alam, Phys. Rev. C , 2009, 80:064906; P. Mohanty, J. K. Nayak, J. Alam et al, arXiv: 0910.4856

\bibitem{ollitrault} J. Y. Ollitrault, Nucl. Phys. A, 1998, 638: 195c

\bibitem{ko-blastwave} Y. Oh, Z. W. Lin,  C. Y. Ko, Phys. Rev. C , 2009, 80: 064902

\bibitem{blastwave} P. Siemens , J.O. Rasmussen, Phys. Rev. Lett., 1979, 42: 880

\bibitem{heinz} P. Huovinen, P.F. Kolb, U. Heinz et al, Phys. Lett. B, 2001, 503: 58

\bibitem{ampt1} B. Zhang, C.M. Ko, B.A. Li et al, Phys. Rev. C
, 2000, 61: 067901

\bibitem{ampt2} Zi-Wei Lin, Che Ming Ko, Bao-An Li et al, Phys. Rev. C, 2005, 72: 064901

\bibitem{ampt-v2} Z.W. Lin and C. M. Ko, Phys. Rev. C , 2002,65: 034904

\bibitem{star-pt-v2} J. Manninen, F. B., Phys. Rev. C , 2008,77: 054901

\bibitem{Rudy} R. C. Hwa, arXiv:1009.0506

\bibitem{Heinz-song} H. Song , U. W. Heinz, Phys.Rev. C, 2008, 77 :064901

\bibitem{Song} Huichao Song, arXiv: 0908.3656


\end{thebibliography}
